\documentclass[prb,aps,showpacs,twocolumn,superscriptaddress]{revtex4}
\usepackage{graphics,bm}
\usepackage{amssymb}
\usepackage{epsfig}
\usepackage{epsf}

\def\be{\begin{equation}} \def\ee{\end{equation}}
\def\bea{\begin{eqnarray}} \def\eea{\end{eqnarray}}
\def\nn{\nonumber}

\begin{document}

\title{Berry Phase Coupling and the Cuprate Neutron Scattering Resonance}

\author{Wei-Cheng Lee}
\affiliation{Department of Physics, University of California, San Diego, CA 92093}
\affiliation{Department of Physics, The University of Texas at Austin, Austin, TX 78712}
\author{A.H. MacDonald}
\affiliation{Department of Physics, The University of Texas at Austin, Austin, TX 78712}

\date{\today}

\begin{abstract}
We examine the influence of coupling between particle-hole and particle-particle spin  
fluctuations on the inelastic neutron scattering resonance (INSR) in cuprate superconductors 
in both weak and strong interaction limits.  For weak-interactions in the particle-hole
channel, we find that the interchannel coupling can eliminate the resonance.  For strong interactions which
drive the system close to a $\bm{Q}=(\pi,\pi)$ magnetic instablity, the resonance frequency always 
approaches zero but its value is influenced by the interchannel coupling.  We comment on constraints 
imposed on cuprate physics by the INSR phenomenology, and a comparison between the cuprates and 
the newly-discovered iron pnictide superconductors is discussed.
\end{abstract}

\maketitle

\section{Introduction} 
The low-temperature properties of cuprate superconductors seem to be well described by 
mean-field-theory with effective interactions which lead to $d$-wave 
superconductivity, except possibly in the extreme underdoped limit.
At the same time the apparently ubiquitous\cite{insr} 
presence of a low-frequency spin-resonance near $\bm{Q}=(\pi,\pi)$ in inelastic neutron scattering experiments
and the appearance of spin order in some systems\cite{lake} when an external magnetic field is applied,   
suggest that superconducting cuprates are close to an antiferromagnetic 
instability.  In previous work\cite{leewc1} we have argued that coupling\cite{Zhang_Demler} 
between particle-hole and particle-particle spin fluctuations likely contributes to the suppression 
of the superfluid density\cite{rhosuppression} observed in underdoped cuprates.
In this article we specifically address, within time-dependent mean-field theory,
the influence of particle-particle to particle-hole coupling
on the inelastic neutron scattering resonance (INSR) position.
When the interchannel coupling is ignored\cite{excitontheories} an arbitrarily
weak particle-hole channel interaction will induce a resonance. 
We find that the interchannel coupling 
can eliminate weak-interaction resonances.  For 
stronger interactions the system can be driven close to 
an antiferromagnetic instability.  In this limit   
we find that the way in which the resonance frequency approaches zero
is influenced by the interchannel coupling.  There is no limit in which
RPA-type theories without the interchannel coupling yield reliable resonance frequency predictions.

\section{Relationship between the diagonal and off-diagonal RPA susceptibilities}
In time-dependent mean-field theory the most general linear response of a 
system consists of correlated particle-hole and particle-particle excitations.
For square-lattice models with on-site and near-neighbor electron-electron interactions
we have shown previously\cite{leewc2} that at ${\bm Q}=(\pi,\pi)$ 
correlations occur between between four different weighted 
particle-hole transition sums.  The most general response function can therefore 
be evaluated by inverting four by four matrices.  For a model (motivated by 
the cuprate superconductivity literature) with only on-site 
($U$) and near-neighbor ($V$) spin-independent interactions and 
nearest-neighbor Heisenberg ($J$) spin interactions we find\cite{leewc2} that:
\begin{equation}
\hat{\chi}^{-1}(\vec{Q},\omega)=\hat{\chi}_{qp}^{-1}(\vec{Q},\omega)-\hat{V}
\label{grpares}
\end{equation}
where $\hat{V}={\rm diag}(-U-2J, J/2-2V,V+J/4,V+J/4)$ 
is the interaction kernel,
\begin{equation}
\begin{array}{l}
\displaystyle
\hat{\chi}_{ab,qp}(\vec{Q},\omega)=\\[2mm]
\displaystyle
\frac{1}{N}\sum_{\vec{k}}\left(\frac{f_a(\vec{k}) f_b(\vec{k})}{\omega-E(\vec{k})-E(\vec{k}')}-
\frac{(-1)^{a+b}f_a(\vec{k}) f_b(\vec{k})}{\omega+E(\vec{k})+E(\vec{k}')}\right)
\end{array}
\label{chi0}
\end{equation}
is the bare mean-field-quasiparticle response function, and $\vec{k}'=\vec{k}+\vec{Q}$ reduced to the first Brillouin-zone.
The indices $a$ and $b$ refers to the four coupled two-particle excitation channels and 
$f_a$ is the corresponding coherence factor defined 
in Ref. [\onlinecite{leewc2}]. As shown in previous work\cite{leewc2}
the INSR mode is dominated by two channels, 
the spin flip $(a=1)$ and the $d$-wave pair phase $(a=4)$ modes,
when parameters are in a range broadly consistent with experiment. 
The same conclusion can be reached on the basis of earlier 
approximate RPA theories\cite{rpa3channel} which omitted the second channel.
The qualitative discussion of INSR physics below is based on a model in which only these two modes are 
retained, although the full 4-channel model is required for quantitative accuracy. 

In the truncated two-channel model the INSR solves 
\be
K^s(\omega)K^\phi(\omega)=C^2(\omega)\omega^2
\label{eres}
\ee
where $K^s(\omega)=K_{qp}^s(\omega)-V_s$, $K^\phi(\omega)=K_{qp}^\phi(\omega)+V_\phi$, $V_s=U+2J$, $V_\phi=V+J/4$, and
the frequency-dependent stiffness and Berry-phase coupling parameters are defined by:
\bea
K_{qp}^s(\omega)&=&-\frac{1/\chi_{11,qp}(\omega)}{1-R(\omega)}\nn\\
K_{qp}^\phi(\omega)&=&-\frac{1/\chi_{44,qp}(\omega)}{1-R(\omega)}\nn\\
C(\omega)&=&\frac{-1}{\omega}\frac{\chi_{14,qp}(\omega)}{\chi_{11,qp}(\omega)\chi_{44,qp}(\omega)(1-R(\omega))}\nn\\
R(\omega)&=&\frac{\chi^2_{14,qp}(\omega)}{\chi_{11,qp}(\omega)\chi_{44,qp}(\omega)}.
\label{cromega}
\eea
$K^s(\omega=0)$ and $K^{\phi}(\omega=0)$ are\cite{leewc1} proportional to the 
energetic cost of static spin-density and spin-polarized electron-pair density
fluctuations, as estimated by mean-field-theory while $C(\omega=0)$ is\cite{leewc1} the 
Berry-curvature associated with adiabatic evolution of electron-electron and 
electron-hole pair mean-fields.  
Stability of the $d$-wave superconducting state requires that 
both $K^s(\omega=0)$ and $K^\phi(\omega=0)$ be positive.  
In the two-channel approximation
the interchannel coupling is captured by the function $\chi_{14,qp}(\omega)$.
If $\chi_{14,qp}(\omega)$ was negligible, Eq. \ref{eres} 
would be satisfied at both the spin-exciton resonance frequency $\omega_{ex}$ defined by  
\be 
\label{eq:omegaex}  
1/\chi_{11,qp}(\omega_{ex})+V_{s}=0,
\ee
and at the $\pi$-resonance frequency $\omega_{\pi}$ defined by 
\be 
1/\chi_{44,qp}(\omega_{\pi})-V_{\phi}=0.
\ee
When coupling between particle-hole and particle-particle channels is neglected both $R(\omega)$ and the Berry phase coupling  
$C(\omega)$\cite{leewc1,leewc2} are set to zero. It is therefore important to study how $R(\omega)$ and $C(\omega)$ depend on the model parameters.
Since both $R$ and $C$ are dimensionless, the unbiased justification of whether or not the interchannel coupling is negligible should be based on these two quantities. 
It is evident from the definition of $R(\omega)$ and from 
the way in which it appears in the above equations, that the interchannel coupling certainly can not be neglected when it approaches $1$.
To judge the importance of the Berry-phase coupling $C(\omega)$, we should compare
$\vert C\vert \times \omega_{res}$ with typical values for $\sqrt{K_{s}\times K_{\phi}}$, which indicates that values of $\vert C\vert$ comparable to $1$ also 
correspond to strong interchannel coupling. 

\begin{figure}
\includegraphics{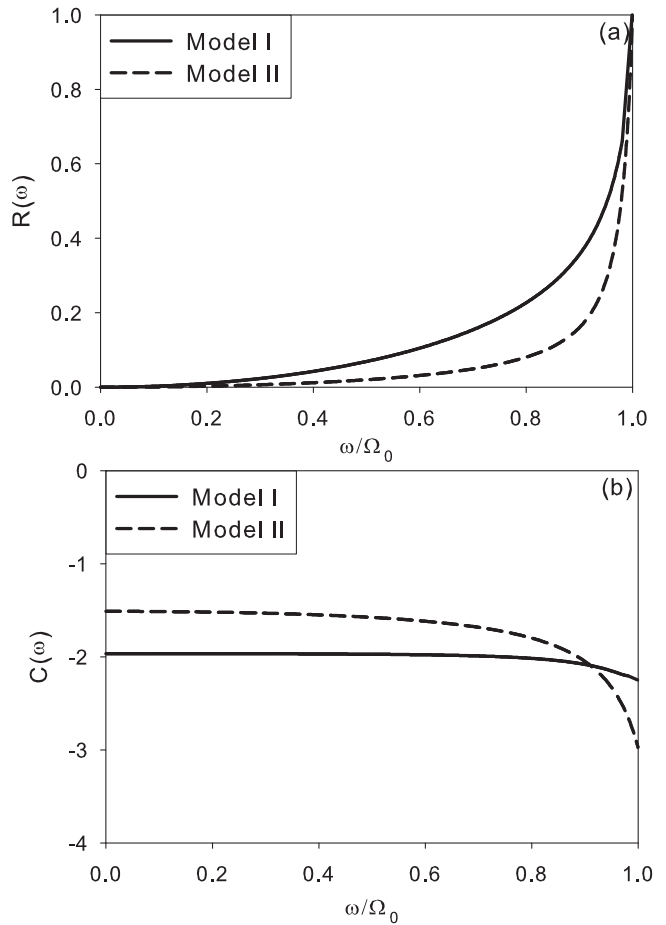}
\caption{\label{fig:ratio} (a) $R(\omega)$ and (b) $C(\omega)$ for two difference band structures. $\Omega_0\sim 70$ meV in both models.
The property that $R(\omega)$ approaches $1$ for $\omega\to\Omega_0$ is a universal feature which applies for any band structure model.
$C(\omega)$ is finite and large in both models.}
\end{figure}

The mean-field-theory of INSR is defined by the hopping parameters of the square-lattice tight-binding model and 
by the interaction parameters $U$,$V$, and $J$. 
We emphasize that all these parameters should be viewed to a large 
degree as effective parameters which depend on microscopic many-body physics in ways which are not fully understood.
Their values are therefore best determined phenomenologically.  Luckily the quantities $R(\omega)$ and $C(\omega)$ depend only on the 
normal state band-structure model, and on the maximum of the mean-field $d$-wave energy gap $\Delta$.
At the current mature state of cuprate experimental research, the information we require 
is therefore available with adequate accuracy. 

Fig. \ref{fig:ratio}(a) plots $R(\omega)$ for $\omega\leq \Omega_0$, where $\Omega_0$ is the minimum two-particle excitation energy, with
two very different band structures. 
Fig. \ref{fig:ratio}(b) plots $C(\omega)$.  
Although the association of $C$ with Berry phases is not strictly appropriate away from the adiabatic limit, 
it is appropriate to retain the terminology because $C(\omega)$ has weak frequency dependence up to quite 
high frequencies. 
Model I in Fig. \ref{fig:ratio} refers to the band structure model obtained by Norman {\it et al.}
\cite{norman} by fitting to angle-resolved-photoemssion data\cite{arpes} and used by us in our previous work\cite{leewc2}.
Model II refers to a model used by Hao and Chubukov\cite{hao} in a related recent work. 
The main difference between these two models 
is that the nearest-neighbor hopping $t$ is almost 2.5 times bigger in Model II than that in Model I.
Since the maximum energy gaps in both models are fitted to be 
around $35-50$ meV in the underdoped regime, the ratio of the energy gap to the kinetic energy is 
much smaller in Model II than that in Model I.  This difference is responsible for the generally  
weaker interchannel coupling effects seen in both the $R(\omega)$ and $C(\omega)$ in Model II as shown in Fig.~[\ref{fig:ratio}].
We note however that $R(\omega) \to 1$ as $\omega$ approaches $\Omega_0$ in 
both models, and indeed as we discuss below, for any band-structure model.
Although Model I seems to be much more sophiscated and better fitted with experiments\cite{norman},
we study both models to ease comparison with Ref. [\onlinecite{hao}] and to 
provide some quantitative indication of the sensitivity of our conclusions to
very-different band-structure model choices. 

\section{Weak Interaction Limit} 
Weak interactions can compete with band energies for $\omega$ smaller 
than but close to the minimum two-particle excitation energy $\Omega_0$ because the 
energy denominator in Eq.~\ref{chi0} becomes small when 
$\bm{k}$ approaches $\bm{p}$, the point at which 
$E(\vec{k})+E(\vec{k'})$ is minimized.   
By expanding the energy denominator around its minimum we find that
all elements of the quasiparticle response function diverge logarithmically
as $\omega \to \Omega_0$ from below:
\bea
\chi_{11,qp}(\omega=\Omega_0^-)&\sim& f^2_1(\vec{p}) \; \nu_0 \;  {\rm ln}(1-\omega/\Omega_0)\nn\\
\chi_{44,qp}(\omega=\Omega_0^-)&\sim& f^2_4(\vec{p}) \; \nu_0 \;  {\rm ln}(1-\omega/\Omega_0)\nn\\
\chi_{14,qp}(\omega=\Omega_0^-)&\sim& f_1(\vec{p})f_4(\vec{p}) \; \nu_0  \; {\rm ln}(1-\omega/\Omega_0).
\label{logdiv}
\eea
Here $\nu_0$ is the joint density of states per site just above the minimum excitation energy. 
Note that all off-diagonal elements of the quasiparticle response matrix approach the geometric mean
of the corresponding diagonal elements, so that the response matrrix becomes 
singular and $R(\omega=\Omega_0^-)\to 1$.  
As we explain below, the familiar 
BCS-like weak-interaction result for the spin-exciton   
resonance frequency is qualitatively altered as a consequence.  

When the interchannel coupling is neglected, the spin exciton energy $\omega_{ex}$ is obtained\cite{excitontheories}
by inserting Eq.~(\ref{logdiv}) in Eq.~(\ref{eq:omegaex}) to obtain
\be
1- \frac{\omega_{ex}}{\Omega_0}  \; = \; \exp\big(-\frac{1}{\nu_0V_sf^2_1(\vec{p})}\big) 
\ee
An undamped resonance appears just below the two-particle continuum edge for 
an infinitesimally small positive $V_s$. To account for the interchannel coupling for $\omega$ 
near $\Omega_0$ we rewrite Eq.~(\ref{eres}) as
\be
(-\chi_{44,qp}(\omega)-D(\omega)V_s)(-\chi_{11,qp}(\omega)+D(\omega)V_\phi)=\chi^2_{14,qp}
\ee
where 
\be
D(\omega)=(\chi_{11,qp}(\omega)\chi_{44,qp}(\omega)-\chi^2_{14,qp}(\omega)).
\ee
Note that, although each matrix element diverges logarithmically the geometric 
product structure of the matrix implies that $D(\omega)$ diverges like $\ln$, not 
like $\ln^2$. After some algebras, we obtain
\be
1-\frac{\omega_{INSR}}{\Omega_0}=\exp\big(\frac{-1}{\nu_0 (V_s f^2_1(\vec{p})-V_\phi f^2_4(\vec{p}))}\big).
\ee
If $V_\phi>0$, $V_s$ must be greater than a critical value in order to have a resonance and 
$\omega_{INSR}>\omega_{ex}$ in agreement with Ref. [\onlinecite{hao}]. 
If $V_\phi\leq 0$, $\omega_{INSR}\leq\omega_{ex}$ and an infinitesimal positive $V_s$ is again
sufficient to guarantee a resonance below the two-particle continuum. 

\section{Strong Interaction Limit} 

Strong interactions drive the system close to the antiferromagnetic instability, {\em i.e.} into the regime which appears to be 
relevant for optimally doped and moderately underdoped cuprates.  Given the instability criterion,
$K^s(\omega=0)=K_{qp}^s(\omega=0)-V_s$ with $V_s=U+2J$ we see that either on-site Coulomb interactions $U$, or Heisenberg  
near-neighbor spin-dependent effective interactions $J$, or a combination of the two, could be responsible.  Since, as we mention again below, 
experiments imply that $K^s(\omega=0)$  is around an order of magnitude smaller than $K_{qp}^s(\omega=0)$, the value of $V_s$ 
is tightly constrained on a relative basis.  It has been argued\cite{hao} that $\omega_{ex}$ should 
match $\omega_{res}$ accurately in a relative sense in this limit because the frequency which appears in the Berry phase coupling 
term is small.  As we will prove in detail later, this argument is not necessarily correct 
because the leading frequency dependent term in the low-frequency expansion of 
Eq.\ref{eres} varies as $\omega^2$, and all $\omega^2$ terms,
including the Berry phase coupling term, should be retained. 

Let's consider the case in which the system is close to the antiferromagnetic instability so that the INSR frequency is small.
We expand the quasiparticle-response dependent quantities at low frequencies as follows:
$\chi_{11,qp}(\omega)\approx -(R^s_0+R^s_2\omega^2)$, $\chi_{44,qp}(\omega)\approx -(R^\phi_0+R^\phi_2\omega^2)$, 
$R(\omega)\approx R_{14}\omega^2$, and $C(\omega)\approx C$.   
Detailed expressions for the positive-definite coefficients $R^{s,\phi}_0$, $R^{s,\phi}_2$ and $R_{14}$ 
can be found in Ref. [\onlinecite{leewc2}]. $R_{14}$ and $C$ are related by
$C= R_{14}/{\cal D}$ where ${\cal D}$ is the low frequency limit of $\chi_{14,qp}(\omega)/\omega$.
The spin-exciton energy can be obtained by setting $V_s \chi_{11,qp}(\omega) = -1$.  We find that 
\be
\omega_{ex}=\sqrt{\frac{1-R^s_0 V_s}{V_s R^s_2}}\equiv \sqrt{\frac{\delta}{V_s R^s_2}},
\ee
where $\delta$ is defined by the second equality.
The above expression is valid close to the antiferromagnetic instability, {\em i.e.} for $\delta = \to 0^+$. 
When the interchannel coupling is included the corresponding resonance energy $\omega_{INSR}$ 
expression is obtained by solving Eq. \ref{eres} retaining terms up to order $\omega^2$.
After some algebras we find:
\be
\label{ga}
\frac{\omega_{INSR}}{\omega_{ex}}=\frac{1}{\sqrt{1+\gamma_1+\gamma_2}}\equiv \gamma
\ee
where
\be
\gamma_1=\frac{D^2}{R^\phi_0 R^s_2}\left(\frac{1}{(1-\delta)(1+V_\phi R^\phi_0)}-1\right)
\label{ga1}
\ee
and
\be
\gamma_2=\frac{-\delta (R^{\phi}_2 - R^\phi_0 R_{14}) V_\phi}{(1+V_\phi R^\phi_0)R_2^sV_s}.
\label{ga2}
\ee
Since $\gamma_2\propto \delta\to 0$ near the spin-density instability,
the discrepancy between $\omega_{INSR}$ and $\omega_{ex}$ is mainly associated with the parameter $\gamma_1$.
We comment further on this point in the next section.
Clearly the value of $\gamma_1$ depends strongly on model parameters. 

\section{Discussion}

As emphasized in our previous work\cite{leewc2}, because we have three 
interaction parameters $(U,V,J)$ but only two experimentally-determined properties, the anti-nodal gap and the INSR frequency, 
experiment does not uniquely determine the effective interaction model.
The two combination of interaction parameter combinations that 
are tightly constrained by experiment are the spin-interaction 
$V_s=U+2J$ (INSR experiment) and the pair-potential $V_{pair}=3J/2 - 2V$
(ARPES and other measures of the $d$-wave antinodal gap).    
We have argued previously that the $V=0, J\ne 0$ ($J$-scenario) possibility
(pairing driven by effective Heisenberg interactions)
is more likely than that $V \ne 0, J=0$ ($V$-scenario) possibility 
(pairing driven by attractive scalar near-neighbor interactions).
Our argument appealed to the {\it Occam's razor} assumption that 
one type of effective interaction is likely to play a dominant role in 
the low-energy physics.  We then observed that  
the $J$-scenario can account for both 
$V_s$ and $V_{pair}$ scales whereas the 
$V$-scenario requires in addition a finely-tuned on-site effective interaction. 
($U$ would have to be at least $350 meV$ in model I and 800meV in model II if the $V$-scenario
was adopted to fit the experimental 
data\cite{leewc2,hao}.)
It is in fact clear from the outset that the near-neighbor scalar interaction
$V$ cannot succeed in accounting for both scales because it contributes to $V_{pair}$ but not to $V_s=U+2J$. 
On the other hand, the $J$-scenario is not trivially guaranteed to explain both experimental energy scales.
This fact does appear to us to be strongly suggestive.

Eqs.~(\ref{ga},\ref{ga1},\ref{ga2}) demonstrate that interchannel coupling can alter the resonance 
frequency even in the limit $K_{s} \to 0$ and that the resonance frequency always depends 
on both $V_{\phi}$ and $V_{s}$.  In the $J$-scenario we favor 
$V_\phi$ is small and positive, and $\gamma_1$ tends to be negative 
in models which are close to the antiferromagnetic instability. 
For this reason $\gamma$ is slightly larger than 1 in both models we have examined.
For Model I, we find $\gamma\approx 1.1$ while for Model II we 
find $\gamma\approx 1.04$.  The approximate coincidence between
$\omega_{ex}$ and $\omega_{INSR}$ for some model parameters does not imply that 
particle-hole and particle-particle modes are weakly coupled in the elementary
excitations. In fact, it is evident in Fig.~(\ref{fig:ratio}) that $C(\omega)$ is weakly frequency 
dependent and remains comparable to $1$ in both models, which already guarantees the strong interchannel coupling as discussed above.
The close match between $\omega_{ex}$ and $\omega_{INSR}$ only reflects that the choice of interaction parameters leads to not only $K^s(\omega)\to 0$ (i.e. $\delta\to 0$) 
but also the large ratio of $K_{\phi}(\omega)/K_{s}(\omega)$ at low frequency. For realistic  
experimental situations in which the resonance frequency is always a substantial fraction
of the antinodal gap, this ratio is always larger than approximately $3$, which does favor $\omega_{INSR}\approx\omega_{ex}$.
This analysis proves that the ratio of $\omega_{INSR}/\omega_{ex}$ can not serve as a reliable criterion to judge the importance of the interchannel coupling.
We have referred to this resonance as a magnetic plasmon to suggest this aspect of 
its character, namely that it represents the coupled quantum fluctuations of 
canonically conjugate degrees of freedom as long as $\vert C\vert$ is comparable to 1. 
Our conclusions on this point are in disagreement with those of 
Ref. [\onlinecite{hao}].

In Ref. [\onlinecite{leewc1}] we have argued that the quantum zero-point energy of 
cuprates makes a negative contribution to the superfluid density of cuprates, and 
that it therefore contributes to the decrease of the critical temperature in the underdoped regime.
The basic idea is that superconductivity is weakened by phase gradients, allowing the 
antiferromagnetic instability to creep closer and reducing the quantum zero-point energy
associated with spin-fluctuations.  Similar physics occurs in 
bilayer quantum Hall superfluids\cite{joglekar,eisenstein}. 
Microscopic caclulations suggest that in cuprates 
this effect is large enough to account for the decreasing superfluid density, in 
contrast to the situation in typical superconductors in which correlation effects 
are largely independent of the configuration of the superfluid condensate.
It will be interesting to examine how this idea plays out when applied 
to the recently-discovered iron pnictide superconductors. Current experiments and theories seem to favor 
extended-s-wave gap symmetry.  An INSR has been observed\cite{christianson,lumsden,chi} in
the pnictide materials with a phenomenology which is surprisingly similar to that of the cuprates.
Because the the ratio $2\Delta/k_B T_c$ is found to be remain close to  
the BCS predcition\cite{chen2008}, it seems that the supression of superfluid density 
by correlations is less important in this material.  
To account for this difference within our theory, we will need to 
understand why phase gradients in the superconducting condensate have a weaker effect 
on spin-fluctuations and on the correlation energy.
The multi-band character and the peculiar Fermi surface pockets of pnictides would seem likely to produce this result.
The large effect in cuprates rests strongly on the importance of fluctuations 
near the wavevector which connects anti-nodal momenta for spin-response functions, 
a coincidence that is not likely to be repeated in iron pnictide superconductors.
A recent RG analysis\cite{chubukovrg} which showed that the particle-hole and particle-particle 
channels are decoupled at energy scalies below the Fermi energy $E_F$ supports this view.
A more detail calculations for the iron pnictides using an appropriate
multiband model which aims to resolve this issue is currently in progress.

Based on these analyses, we conclude that Berry phase coupling between the particle-hole and particle-particle channels can never be ignored in any RPA-type theory of the 
inelastic neutron scattering resonance mode for cuprates.
As a result, the inelastic neutron scattering resonance mode observed in cuprates should be bettter interpreted as a magnetic plasmon than a spin exciton.
The importance of coupling in the pnictides has not yet been adequately analyzed.
Nevertheless, the spin resonance mode observed in the inelastic 
neutron scattering measurements could be a purer spin exciton, and the 
magnetic and superconducting instablities less coupled in the pnictide case.

This work is supported by the US Army Research Office under grant No. ARO-W911NF0810291 and the National Science Foundation under grant No. DMR-0606489. 
We would like to thank A.A. Burkov, A. Chubukov, Y. Joglekar, Z. Hao, J. Sinova, and C. Wu for helpful discussions.


\begin{thebibliography}{99}
\bibitem{insr} J. Rossat-Mignon {\em et al.}, Physica. C {\bf 185}, 86 (1991); H.
Mook {\em et al.}, Phys. Rev. Lett. {\bf 70}, 3490 (1993); H.F. Fong
{\em et al.}, Nature {\bf 398}, 588 (1999); H. He {\em et al.}, Science {\bf 295},
1045 (2002); G. Yu et al, to appear.
\bibitem{lake} B. Lake {\em et al.}, Nature {\bf 415}, 299 (2002); J. Chang {\em et al.}
arXiv:cond-mat/0712.2182; Ying Zhang, Eugene Demler,and Subir Sachdev, Phys. Rev. B {\bf 66}, 094501 (2002).
\bibitem{leewc1} Wei-Cheng Lee, Jairo Sinova, A.A. Burkov, Yogesh Joglekar, and A.H. MacDonald, Phys. Rev. B. {\bf 77}, 214518 (2008).
\bibitem{Zhang_Demler} E. Demler, H. Kohno, and S. C. Zhang, Phys. Rev. B {\bf 58}, 5719 (1998) and work cited therein.
\bibitem{rhosuppression} Y.J. Uemura {\em et al.}, Phys. Rev. Lett. {\bf 62}, 2317 (1989); V.J. Emery and S.A. Kivelson, Nature {\bf 374}, 434 (1995).
\bibitem{excitontheories} M. Eschrig, Adv. Phys. {\bf 55}, 47 (2006) and references therein.
\bibitem{leewc2} Wei-Cheng Lee and A.H. MacDonald, Phys. Rev. B. {\bf 78}, 174506 (2008).
\bibitem{rpa3channel} O. Tchernyshyov, M. R. Norman, and A. V. Chubukov, Phys. Rev. B 63, 144507 (2001).
\bibitem{arpes} A. Damascelli, Z. Hussain, and Z.X. Shen, Rev.Mod. Phys. {\bf 75}, 473 (2003).
\bibitem{hao} Zhihao Hao and A.V. Chubukov, arXiv:0812.2697 (2008).
\bibitem{norman} M.R. Norman, M. Randeria, H. Ding, and J.C. Campuzano, Phys. Rev. B {\bf 52}, 615 (1995).
\bibitem{joglekar} Yogesh N. Joglekar and Allan H. MacDonald , Phys. Rev. B {\bf 64}, 155315 (2001).
\bibitem{eisenstein} J.P. Eisenstein and A.H. MacDonald, Nature {\bf 432}, 691 (2004).
\bibitem{christianson} A. D. Christianson, E. A. Goremychkin, R. Osborn, S. Rosenkranz, M. D. Lumsden, C. D. Malliakas, I. S. Todorov, H. Claus, 
D. Y. Chung, M. G. Kanatzidis, R. I. Bewley, T. Guidi, Nature {\bf 456}, 930 (2008).
\bibitem{lumsden} M. D. Lumsden, A. D. Christianson, D. Parshall, M. B. Stone, S. E. Nagler, G. J. MacDougall, H. A. Mook, K. Lokshin, T. Egami, D. L. Abernathy, 
E. A. Goremychkin, R. Osborn, M. A. McGuire, A. S. Sefat, R. Jin, B. C. Sales, and D. Mandrus, Phys. Rev. Lett. {\bf 102}, 107005 (2009).
\bibitem{chi} Songxue Chi, Astrid Schneidewind, Jun Zhao, Leland W. Harriger, Linjun Li, Yongkang Luo, Guanghan Cao, Zhu'an Xu, Micheal Loewenhaupt, Jiangping Hu, 
and Pengcheng Dai, Phys. Rev. Lett. {\bf 102}, 107006 (2009).
\bibitem{chen2008} T. Y. Chen, Z. Tesanovic, R. H. Liu, X. H. Chen, C. L. Chien, Nature {\bf 453}, 1224 (2008).
\bibitem{chubukovrg} A.V. Chubukov, arXiv:0902.4188 (2009).
\end{thebibliography}
\end{document}